\documentclass[a4paper,fleqn,usenatbib]{mnras}
\usepackage[T1]{fontenc}
\usepackage{ae,aecompl}

\usepackage{graphicx}

\usepackage{url}
\usepackage{amsmath}
\usepackage{amssymb}
\newcommand{\eq}[1]{Equation~\ref{eq:#1}}
\newcommand{\tite}{$\frac{U_{e}}{U_{sim}}$}
\newcommand{\titel}{$U_{e}/U_{sim}$}

\title{The impact of non-thermal electrons on event horizon scale images and spectra of Sgr A*}

\author[Mao, Dexter, \& Quataert]{
S. Alwin Mao,$^{1}$
Jason Dexter,$^{2}$
and Eliot Quataert$^{3}$
\\
$^{1}$Department of Astrophysics, Princeton University, Princeton, NJ 08544\\
$^{2}$Max Planck Institute for Extraterrestrial Physics, Giessenbachstr. 1, 85748 Garching, Germany \\
$^{3}$Astronomy and Physics Departments and Theoretical Astrophysics Center, \
University of California, Berkeley, CA 94720
}

\pubyear{2016}
\begin{document}

\label{firstpage}
\pagerange{\pageref{firstpage}--\pageref{lastpage}}
\maketitle

%
%

\begin{abstract}
\textnormal{Decomposing an arbitrary electron energy distribution into sums of Maxwellian and power law components is an efficient method to calculate synchrotron emission and absorption. We use this method to study the effect of non-thermal electrons on submm images and spectra of the Galactic center black hole, Sgr A*. We assume a spatially uniform functional form for the electron distribution function and use a semi-analytic radiatively inefficient accretion flow and a 2D general relativistic MHD snapshot as example models of the underlying accretion flow structure. }
We develop simple \textnormal{analytic models} which allow us to generalize from the numerical examples. A high energy electron component containing a small fraction (few per cent) of the total internal energy (e.g. a ``power law tail'') can produce a diffuse halo of emission, which modifies the observed image size and structure. 
A population of hot electrons with a larger energy fraction (e.g. resulting from a diffusion in electron energy space) can dominate the emission, 
so that the observed images and spectra are well approximated by considering only a single thermal component for a suitable choice of the electron temperature. We discuss the implications of these results for estimating accretion flow or black hole parameters from images and spectra, and for the identification of the black hole ``shadow'' in future mm-VLBI data. In particular, the location of the first minimum in visibility profiles does not necessarily correspond to the shadow size as sometimes assumed. 
\end{abstract}


\section{Introduction}

Using Very Long Baseline Interferometry (VLBI), the Event Horizon Telescope project (EHT) is, for the first time, attaining radio images which resolve event horizon scale structure around black holes. Such observations could detect the ``shadow'' cast by the black hole on the emission from surrounding gas \citep{bardeen1973,falcke}, a signature of strong field gravity. The current objects of interest are the black holes with the largest angular sizes: supermassive black holes at the centers of nearby galaxies. In particular, Sagittarius A* (Sgr A*) at the center of the Milky Way, has the largest apparent size, mostly due to its proximity \citep[$M_{\rm BH} \simeq 4\times10^6 M_\odot$ at a distance $R_0 \simeq 8$ kpc.][]{chatzopoulosetal2015}.

The observed image depends on the properties of the emitting particles near the black hole. The primary source of radio emission from the environment around Sgr A* is electron synchrotron radiation, and the nature of this emission depends on the energy distribution of the emitting electrons. Both the normalization and shape of the electron energy distribution are uncertain. Sgr A* radiates at a very small fraction ($\sim 10^{-8}$) of its Eddington luminosity, both due to a low accretion rate  \citep{aitken2000,agol2000,quataert2000,bower03,marrone2007} and likely a low radiative efficiency \citep[e.g.,][]{yuan2003}. The low accretion rate implies low particle densities and large mean free paths in the accretion flow. Since the Coulomb collision timescale is much longer than the local dynamical time, it is expected that the electrons and ions do not remain thermally well coupled and there is no reason for the electron distribution function to be Maxwellian. Indeed, a small population of electrons accelerated to higher energies could explain the flat observed radio spectrum \citep{ozel2000,yuan2003}. 

Models of Sgr A* from MHD simulations often assume purely thermal distributions of electrons with a temperature set as a constant fraction of the ion temperature \citep{goldston2005,monika2009,dexter2010}, as a function of fluid properties such as magnetic field strength \citep{moscibrodzkafalcke2013}, or 
by solving a separate electron energy equation \citep{sharma2007,shcherbakovetal2012,ressler2015}. 
Alternatively, semi-analytic models have also used a hybrid thermal and power law distribution function \citep{ozel2000,yuan2003,broderickloeb2006}. \textnormal{More recently, \cite{ball2016} considered injection of non-thermal electrons based on the local magnetic field strength.} From fitting spectra and images with such models, one can obtain model-dependent estimates for parameters of the accretion flow and black hole, such as the accretion rate or black hole spin \citep[e.g.,][]{bro2009b,dexter2010}.

Here we study the effects of a non-thermal distribution function on predicted sub-millimeter images of Sgr A*. We use relativistic radiative transfer calculations (\S\ref{sec:grtrans}) using two sample distribution functions (\S\ref{sec:edf}) and models for the accreting gas (\S \ref{sec:fluid}) to calculate sample images. We show that low energy electrons (below the brightness temperature for an event horizon scale angular size) do not necessarily contribute at all to the observed spectrum and image, leading to large uncertainties in accretion rate estimates from fitting models to data. In addition, even a small population of high energy electrons can lead to an extended, low surface brightness ``halo'' of emission (\S\ref{sec:results}). The presence of this halo affects the inferred image size, and can modify or remove the null in the visibility amplitude caused by the presence of the black hole shadow. The image size can be understood analytically (\S \ref{sec:eis}) in terms of the overall fraction of the total thermal energy density in the highest energy electrons  (\S\ref{sec:emeta}), and so should apply to a wide range of possible distribution functions beyond the two specific examples studied here. 

\section{GRTRANS}\label{sec:grtrans}

We simulate EHT images using the public relativistic radiative transfer code \texttt{grtrans}\footnote{\url{https://github.com/jadexter/grtrans}} \citep{dexter2016}, which uses the public code \texttt{geokerr} \citep{geokerr2009,geokerr2010} to perform ray tracing from an observer's camera to a simulated black hole and its accretion flow. Each pixel on the camera corresponds to a ray. Each point along each ray is assigned values describing the fluid, such as particle number densities, gas pressures, temperatures, and magnetic field strengths, according to a fluid model (see \S\ref{sec:fluid}). Then, \texttt{grtrans} solves the radiative transfer equation along each ray with those fluid values, assuming that the primary emission and absorption are electron synchrotron, and using a prescription for the electron energy distribution function (see \S\ref{sec:edf}). The result is a 2-D image of the accretion flow, as seen from a distant observer. A Fourier transform of the image provides a visibility, simulating an observable from VLBI (in particular, the EHT). The image can also be summed, and multiple images at multiple frequencies can simulate a spectrum.  

We use an image resolution of $150 \times 150$ pixels (rays) covering $50 R_{g} \times 50 R_{g}$ with 400 points on each pixel (ray), which numerically converges with an accuracy at the few per cent level \citep{dexter2010,dexter2016}. \textnormal{We pad the images with zeros to a size of $1024 \times 1024$ before performing Fourier transforms to improve the resolution in Fourier space.} The simulation units are set by the black hole mass. For a given black hole spin, inclination, and electron energy model, we adjust $\dot{M}$ (or equivalently, the normalization of the particle number density in our fluid model) in order to match the observed 230 GHz (1.3 mm) flux, $\approx 3.4$ Jy, following the approach of \cite{monika2009}. Further constraints come from the spectral shape in the radio and infrared, the spectral index and angular size in the submm. We consider these constraints when determining the allowed range of distribution functions (e.g. the fraction of internal energy in non-thermal electrons).

\section{Electron Models}\label{sec:edf}

We calculate images and spectra using three sample electron distribution functions: one thermal distribution, and two physically-motivated extensions of this model to non-thermal distributions. \textnormal{For this work, we assume that the functional form of the electron distribution function is spatially uniform. Other complementary work \citep[e.g.,][]{ball2016} assumes that the energy fraction of the non-thermal electrons spatially varies with the 
strength of the magnetic field. This may arise from electron acceleration in magnetic reconnection or shock regions. We instead consider uniform injection as has been widely done for RIAF models \citep[e.g.,][]{ozel2000,yuan2003} and is necessary to explain the radio spectrum of Sgr A* within these models. This simplification allows us to analytically understand how image properties connect to the electron distribution function (\S\ref{sec:eis}), while making predictions for the images and spectra arising from a spatially uniform electron distribution function. 
Because the electron synchrotron cooling time is long at the sub-millimeter wavelengths considered here, the non-thermal electrons are much more likely to be volume filling than electrons emitting at shorter wavelengths (e.g. infrared). 
}

\subsection{Thermal distribution}

For a gas of relativistic particles, the thermal Maxwellian distribution is described by:
\begin{equation}\label{eq:pltfth}
f(\gamma) = \frac{\gamma^{2}\beta}{\theta K_{2} (1/\theta)}\exp(-\frac{\gamma}{\theta}),
\end{equation}
where the electron Lorentz factor $\gamma = 1/\sqrt{1 - v^{2}/c^{2}}$, $\beta = \frac{v}{c} = \sqrt{1 - 1/\gamma^{2}}$, the dimensionless electron temperature $\theta=\frac{k_{B}T}{m_{e}c^{2}}$, and $K_{2}$ is the modified Bessel function of the second kind. Given a thermal electron number density $N_{th}$, the number density with Lorentz factor $\gamma$ is 
\begin{equation}\label{eq:pltnth}
n_{th}(\gamma) = N_{th}\frac{\gamma^{2}\beta}{\theta K_{2} (1/\theta)}\exp(-\frac{\gamma}{\theta}).
\end{equation}
This has total energy density
\begin{equation}\label{eq:pltuth}
u_{th} = a(\theta)N_{th}m_{e}c^{2}\theta
\end{equation}
where
\begin{equation}\label{eq:plta}
a(\theta) = \frac{1}{\theta} \left[ \frac{3K_{3}(1/\theta) + K_{1}(1/\theta)}{4K_{2}(1/\theta)} - 1 \right]
\end{equation}
is between 3/2 for non-relativistic electrons and 3 for fully relativistic electrons. 

\subsection{Power-law Tail}\label{sec:plt}

\begin{figure}
\includegraphics[width=\columnwidth]{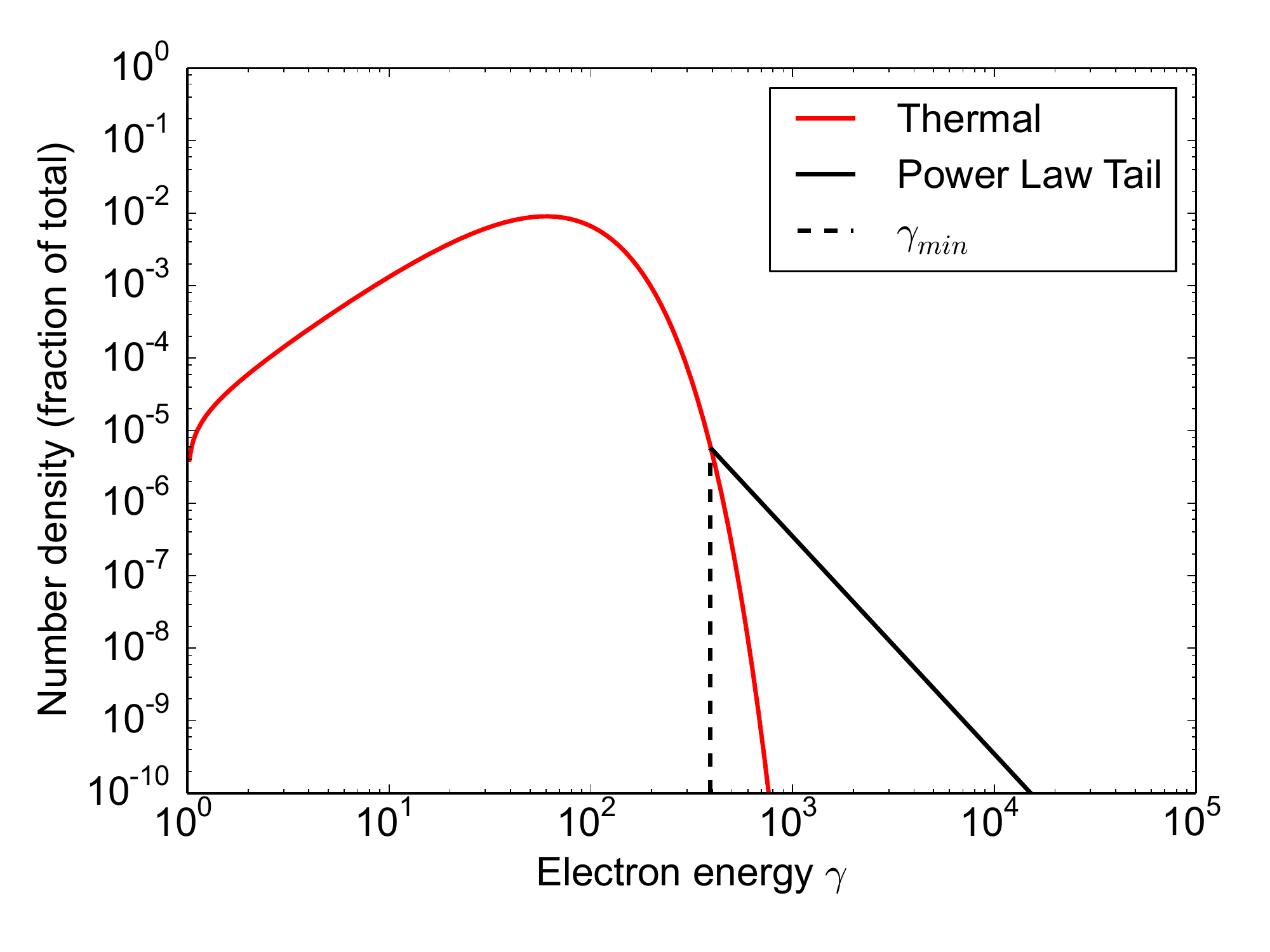}
\caption{An example of a hybrid population of electrons as described in \S\ref{sec:plt} with an electron temperature of $\theta_{e} = 30$, non-thermal energy fraction $\eta = 0.01$, and power $p = 3$. The power law tail begins at $\gamma \simeq 300$, $\simeq 10$ times the electron temperature.}
\label{fig:hybridex}
\end{figure}

The first non-thermal electron energy distribution we investigate is a Maxwellian with a power-law tail of high energy electrons, following \citet{ozel2000} and \citet{yuan2003}. The physical picture is that the electrons are approximately thermal, but a small fraction are accelerated into a higher energy power-law tail. 

To stitch a power-law tail on to the thermal distribution above, we define 
\begin{equation}\label{eq:pltnpl}
n_{pl}(\gamma) = N_{pl} (p-1) \gamma^{-p },~\gamma_{min} \leq \gamma,
\end{equation}
where $\gamma_{min}$ is the Lorentz factor at which the thermal distribution becomes a power-law tail. This distribution has energy density
\begin{equation}\label{eq:pltupl}
u_{pl} = N_{pl}m_{e}c^{2} \frac{p-1}{p-2} \gamma_{min}^{2-p}. 
\end{equation}

Given a power-law index p, it remains to determine $N_{pl}$ and $\gamma_{min}$. In order to solve for these 2 variables, we enforce 2 conditions: continuity at the interface so that $n_{th} (\gamma_{min}) = n_{pl} (\gamma_{min})$, and that the power-law tail contains a fraction of the thermal energy $\eta = \frac{u_{pl}}{u_{th}}$. We leave $\eta$ as a free parameter in order to study how images change based on the amount of energy allotted to the non-thermal electrons. Enforcing the two constraints, we find an equation relating $\gamma_{min}$ and $\theta$ in terms of the parameters $\eta$ and $p$:

\begin{equation}\label{eq:pltgmin2}
a(\theta)(p-2)\eta = \frac{\gamma_{min}^{4}\beta\exp(-\gamma_{min}/\theta)}{\theta^{2}K_{2}(1/\theta)}.
\end{equation}

\noindent In the ultra-relativistic case considered here, $a(\theta) \simeq 3$, $\beta \simeq 1$, and $K_{2} (1/\theta) \simeq 2 \theta^2$ so that this becomes

\begin{equation}\label{eq:pltbessel}
6(p-2)\eta \approx \left(\frac{\gamma_{min}}{\theta}\right)^{4} \exp(-\gamma_{min}/\theta).
\end{equation}

Choosing $p$ and $\eta$, we can solve this equation for the constant $\gamma_{min}/\theta$, which provides a simple linear relationship between $\gamma_{min}$ and $\theta$. Moreover, the exponential behavior of $\exp(-\gamma_{min}/\theta)$ causes the constant $\gamma_{min}/\theta$ to vary slowly for a reasonable range $p$ and $\eta$. For $p = 3.5$, the primary power-law tail we use, varying $\eta$ 14 orders of magnitude between $0.5 \times 10^{-14}$ and 0.5 results in only 1 order of magnitude change in $\gamma_{min}$, from 46 to 4.6. Only the form $(p-2) \eta$ enters in equation \ref{eq:pltbessel}, so that results for the submm image at 230 GHz are also largely insensitive to the shape of the spectral tail (value of $p$). In particular, in the mm, electrons at energies $\sim \gamma_{min}$ are the most important, while $p$ primarily affects the spectrum in the infrared. 

In practice, \eq{pltgmin2} is numerically solvable for a variety of $\eta$, $p$, and $\theta$, and fairly simple fitting functions with tabulated coefficients can be used to precisely approximate $\gamma_{min}$ to within $4\%$. The largest errors only occur for small (nonrelativistic) values of $\theta$, which are irrelevant in a largely relativistic fluid. We used the more accurate approach, solving \eq{pltgmin2} rather than \eq{pltbessel}. 

Either of these approximate numerical techniques allow us to determine $N_{pl}$ and $\gamma_{min}$ at very low computational cost. A synchrotron emission model then determines the emission and absorption coefficients $j_{\nu}$ and $\alpha_{\nu}$, which is dependent on these power-law parameters ($\gamma_{min}$, $N_{pl}$, and $p$). An example of a hybrid thermal and power law tail electron energy distribution is depicted in Figure \ref{fig:hybridex}.

\begin{figure}
\includegraphics[width=\columnwidth]{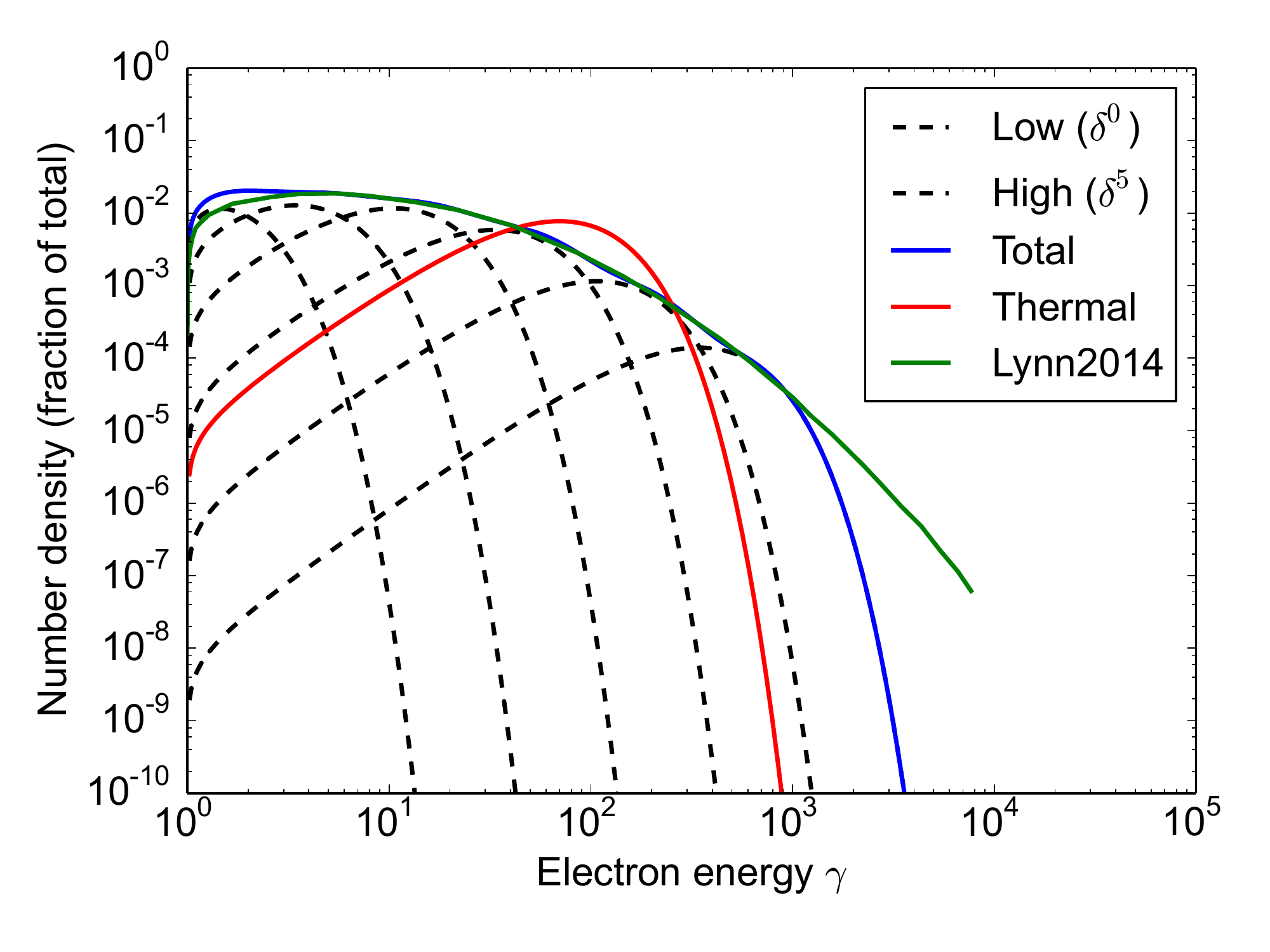}
\caption{An example of the multi-Maxwellian model described in \S\ref{sec:dm} for a population of electrons with electron temperature $\protect\theta_{e} = 35$ and spacing $\protect\delta = 3$. The component Maxwellians are shown in black dashed lines, and the total distribution function is shown in a blue solid line. Comparing the total multi-Maxwellian distribution to an equivalent thermal distribution (red solid line) with the same total electron energy and total number density, the total multi-Maxwellian distribution has more high energy electrons, compensating by also including more low energy electrons. The multi-Maxwellian distribution was produced by a fit to the curve taken from \protect\cite{lynn2014} as shown in green.
}\label{fig:maxexa}
\end{figure}

\subsection{Multi-Maxwellian Model}\label{sec:dm}

In order to extend our study of non-thermal electrons to other electron energy distributions, it would be very useful to have a general technique for implementing an arbitrary electron energy distribution. Integrating over the electron energy distribution at all points along each geodesic would be the most general approach, \textnormal{but would require roughly 10 times the expense of our method for the same accuracy, assuming $\simeq 100$ energy bins for direct integration and $\simeq 6-10$ with component fits. The decomposition further allows us to calculate images and spectra separately for the various components, which we use in \S \ref{sec:eis} to generalize the results here to arbitrary distribution functions and fluid models.}

Instead, it would be possible either to i) find fitting functions or calculate lookup tables for a desired non-thermal distribution function for desired parameter ranges, or ii) describe non-thermal distribution functions as linear combinations of thermal and power-law distributions, for which we have good approximations for the emission and absorption coefficients \citep[e.g.,][]{maha,yuan2003,leungetal2011}. We find the second method more convenient to implement and so use it in this work. For example, the power-law tail distribution function combines a single Maxwellian distribution with a single power-law tail, and then finds the emission and absorption coefficients for the two distributions separately. Then, the net emission and absorption is calculated by combining the coefficients for each distribution. 

As another example of this strategy, we decompose the distribution function of \cite{lynn2014} (see Figure 4 therein) into a number of Maxwellians (multi-Maxwellian). \cite{lynn2014} used magnetohydrodynamic (MHD) simulations with charged particles to calculate the effect of MHD turbulence
on the electron distribution through particle acceleration and momentum-space diffusion. This is a form of Fermi acceleration due to MHD turbulence. In general, black hole accretion flows, such as Sgr A*, are believed to be MHD turbulent through the magnetorotational instability \citep{balbushawley1998}. We approximate the results from \cite{lynn2014} for Fermi acceleration by choosing a number $N$ of Maxwellians to use, which determines the resolution of the decomposition. This results in having to determine $N$ densities $n_{i}$ and $N$ temperatures $\theta_{i}$. We simplify the model so that the Maxwellians are equally spaced in log space, so that $\theta_{i} = \theta_{0}\delta^{i}$, where $\delta$ is the spacing. Each Maxwellian $f_{i}(\gamma)$ is described by 
\begin{equation}\label{eq:dmmax}
f_{i}(\gamma) = n_{i}\frac{\gamma^{2}\beta}{\theta_{i}K_{2} (1/\theta_{i})}\exp(-\frac{\gamma}{\theta_{i}})
\end{equation}
and has total number density $n_{i}$ and total energy density given by Equation~\ref{eq:pltuth}.

Note that the total number density of electrons $n$ should be equal to the sum of the number densities in each Maxwellian. Hence, $n = \sum_{0}^{N-1} n_{i}$, which makes it convenient to define $c_{i} = \frac{n_{i}}{n}$ where $\sum_{0}^{N-1} c_{i} = 1$. These normalized constants determine the shape of the electron energy distribution. Throughout this work we use the coefficients $\{c_{i}\} = \{0.017, 0.074, 0.225, 0.364, 0.228, 0.088\}$
, which fit the electron distribution function from second-order Fermi acceleration at 2 different times (Figure 4 of \cite{lynn2014}) by changing $\delta$ from $\delta \approx 2$ to $\delta \approx 3$. An example of this distribution is shown in Figure \ref{fig:maxexa}. The only case where we use a different set of coefficients is in Figure \ref{fig:emeta}, where ``1/10 Maxwellian'' decreases $c_{5}$ by a factor of 10 and renormalizes all of the coefficients. 

In order to apply this model to a fluid with a given electron number density $n$ and electron temperature $\theta$, having already determined $c_{i} = \frac{n_{i}}{n}$ and $\delta$ from the shape of the desired energy distribution, first note that this temperature $\theta$ implies an energy density $u = a(\theta) n k_{B} T= a(\theta) n \theta m_{e}c^{2}$. By demanding, for consistency, that the total energy density in the $N$ Maxwellians is equal to the given fluid energy density, we require
\begin{equation}
\begin{split}
a(\theta)\theta &= \displaystyle\sum_{i=0}^{N-1} a(\theta_{i})\frac{n_{i}}{n}\theta_{0}\delta^{i}. \label{eq:dmsolve}
\end{split}
\end{equation}

In the ultrarelativistic regime, we can approximate $a(\theta) = a(\theta_{i}) = 3$, so that
\begin{equation}\label{eq:dmdwsum}
\theta_{0} = \frac{\theta}{\displaystyle\sum_{i=0}^{N-1} c_{i}\delta^{i}}. 
\end{equation}

\noindent Hence, given $n$ and $\theta$ from a fluid model, and selecting $c_{i}$, $N$, and $\delta$ for the shape of the distribution, we can solve for $n_{i} = c_{i} n$ and $\theta_{i} = \theta_{0} \delta^{i}$ and then calculate the emission and absorption coefficients for each component in the distribution function.

This method leads to a factor of $< N$ slowdown compared to using a thermal distribution, from $N$ calls to the thermal synchrotron radiation solver. Here $N=6$, so this multi-Maxwellian approach is a computationally feasible method of implementing arbitrary non-thermal electron energy distribution models.

\section{Fluid Models}\label{sec:fluid}

Calculating model images and spectra using the distribution functions described above requires models for the state of the accreting gas near Sgr A*. We use two radiatively inefficient accretion flow (RIAF) models \citep[e.g.,][]{narayanetal1995,yuan2003} as example fluid models. One is a numerical solution of the general relativistic ideal MHD equations, while the other is a self-similar version of the solution found by \citet{yuan2003}.

\subsection{HARM}\label{sec:harm}

We use a single snapshot of a general relativistic MHD solution for accretion onto a spinning black hole in axisymmetry with a spin of $a=0.9375$, calculated using the public code \textsc{HARM} \citep{gammie2003,noble2006}. The code evolves the ideal MHD equations in a Kerr spacetime starting from a torus of gas in hydrostatic equilibrium threaded with a weak magnetic field. The initial configuration is unstable to the magnetorotational instability (MRI), which leads to turbulence and stresses in the fluid, causing angular momentum transport outward and accretion onto the central black hole. The sample snapshot used here is from \citet{dexter2010}, and is from a time $t = 2000 GM/c^3$, once turbulence has developed in the gas but before it dies down due to the axisymmetric simulation grid. 

We scale the dimensionless code units to cgs units as in previous work \citep[e.g.,][]{schnittman2006,noble2007,dexter2010} by fixing the black hole mass and accretion rate. The accretion rate is chosen by normalizing the $230$ GHz flux of the model to $3.4$ Jy. The electron internal energy is set to a constant fraction of the total internal energy in the simulation (from electrons and ions). We denote this fraction as \tite{}. For a thermal distribution function, this is equivalent to choosing a fixed ion-electron temperature ratio ($T_i/T_e$).

Although we consider only a single simulation snapshot instead of averaging observables over time \citep[e.g., as done in][]{monika2009}, we find similar results for submm spectral indices and image sizes when using a thermal distribution function. \textnormal{This is not surprising, since for images and static spectra, 2D HARM simulations have been shown to agree well with time-averaged, 3D calculations \citep{dexter2010} for these types of models. This is in contrast to models based on tilted disks \citep{dexterfragile2013} or unstable jets \citep[][]{medeiros2016} which exhibit a large degree of structural variability. The simulation used here is a reasonable example for our purpose of studying the effects of varying the electron distribution function. We use these fluid examples to test our analytic results described in \S\ref{sec:eis}.}

\subsection{SARIAF}\label{sec:sariaf}

The semi-analytic radiatively inefficient accretion flow fluid model (SARIAF) is a self-similar approximation \citep{bro2009b} to a 1D hydrodynamics accretion solution including electron heating and mass loss \citep{yuan2003}, where fluid variable profiles are as follows:
\noindent the number density of thermal electrons,
\begin{equation}\label{eq:sariafn}
n_{e,th} = n^{0}_{e,th} \left(\frac{r}{r_{s}}\right)^{-1.1} e^{-z^{2}/2\rho^{2}}
\end{equation}

\noindent the temperature of thermal electrons,
\begin{equation}\label{eq:sariaft}
T_{e} = T^{0}_{e} \left(\frac{r}{r_{s}}\right)^{-0.84} 
\end{equation}

\noindent and a toroidal magnetic field in approximate ($\beta = 10$) equipartition with the ions,
\begin{equation}\label{eq:sariafb}
\frac{B^{2}}{8\pi} = \beta^{-1} n_{e,th} \frac{m_{p}c^{2}r_{s}}{12 r}
\end{equation}
where $r_{s} = 2GM/c^{2}$ is the Schwarzschild radius, $\rho$ is the cylindrical radius, and $z$ is the vertical coordinate. 

For modeling Sgr A*, we choose $\beta = 10$ as a fiducial value and vary $n^{0}_{e,th}$ for a chosen $T^{0}_{e}$ to fix the $230$ GHz flux at $3.4$ Jy in the same way as choosing $\dot{M}$ in the HARM model described above. Total intensity simulated images of Sgr A* using this model are in good agreement with those in \citet{broderickloeb2006} for similar parameters.

The radial dependence reflected in this model provides a convenient semianalytic means to understand how different radii contribute to the overall size of an image of the accretion flow, or to the shape of a model spectrum. This is most apparent in Section \ref{sec:sas}, where we use the radial dependence in a SARIAF model to predict the size of an image. Furthermore, unlike the HARM simulation data, which has a defined boundary, the SARIAF model is self-similar and so can be used at all radii. This is especially useful in our study of the size of EHT images, to ensure that the sizes we find from our simulations are not biased by artificial simulation boundaries.

\normalfont

\begin{figure*}
\includegraphics[width=\textwidth]{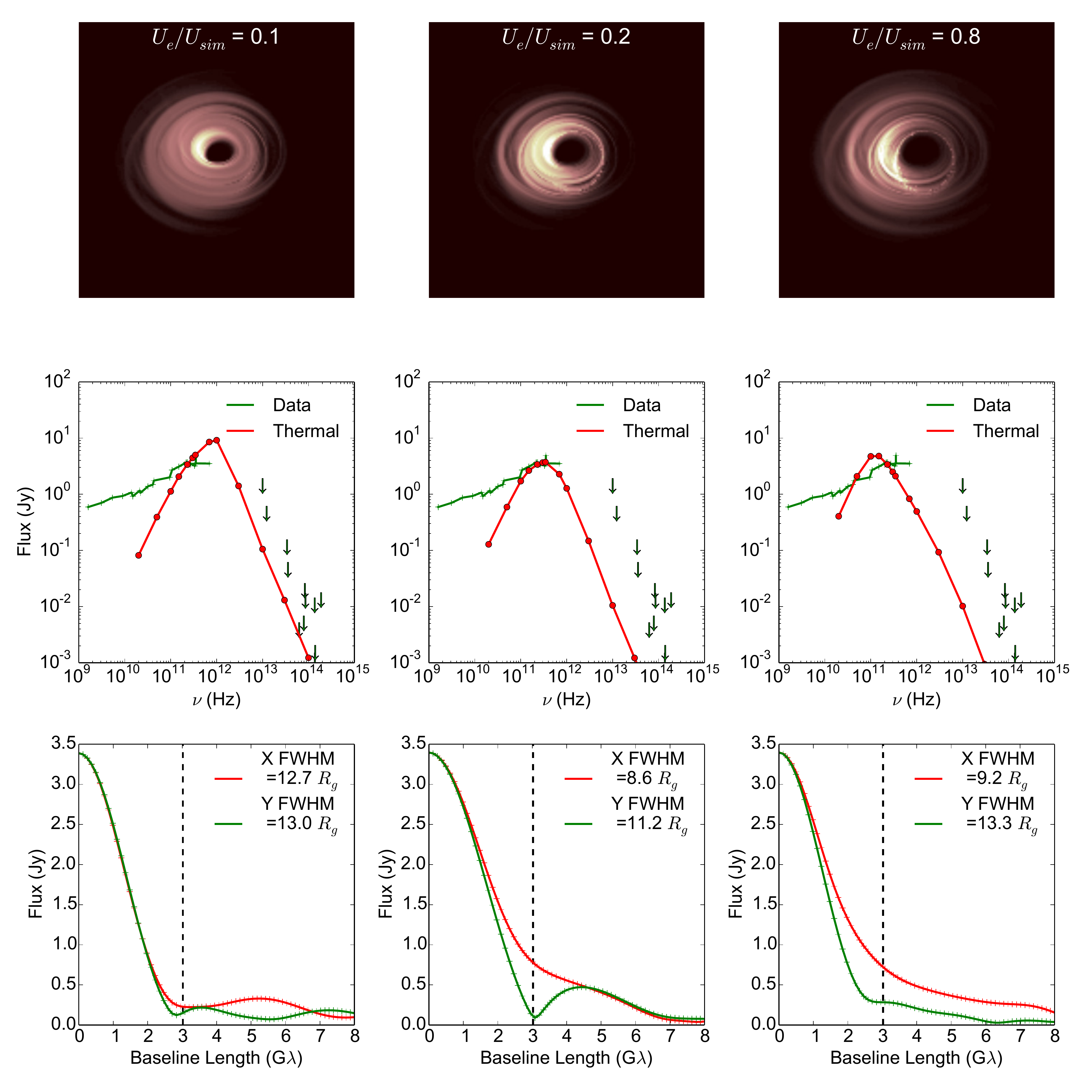}
\caption{A sequence of purely thermal images (top), spectra (middle), and visibility profiles (bottom) from a HARM fluid model of Sgr A* illustrating the effect of increasing electron temperature (left to right)  at fixed 230 GHz flux density and inclination cos i = 0.7. The spectra (red lines and circles) are compared to Sgr A* data (green lines and upper limits as arrows) in the radio \citep{falcke1998,an2005}, submm \citep{marronephd,boweretal2015}, and infrared \citep{meliafalcke2001,genzel2003,doddsedenetal2011}. Note that the image size minimum is at an intermediate temperature and that the spectral peak shifts to lower frequencies at higher temperatures. Images depict a 50 x 50 $R_{g}$ (or 25 x 25 $R_{s}$) region. The red and green visibility profiles are for the X and Y axis in the image, respectively, representing the two extreme perpendicular axes. The full-width-half-maximum size in real space is estimated by fitting a Gaussian to the visibility profile (Fourier space), inverting the size to the real space size, and multiplying by $2\sqrt{2\ln{2}}$ for the FWHM real space size. The vertical dashed line represents the predicted location ($\sim 3 G\lambda$) of a minimum in the visibility profile due to the black hole shadow. 
}\label{fig:thermalseq}
\end{figure*}

\begin{figure*}
\includegraphics[width=\textwidth]{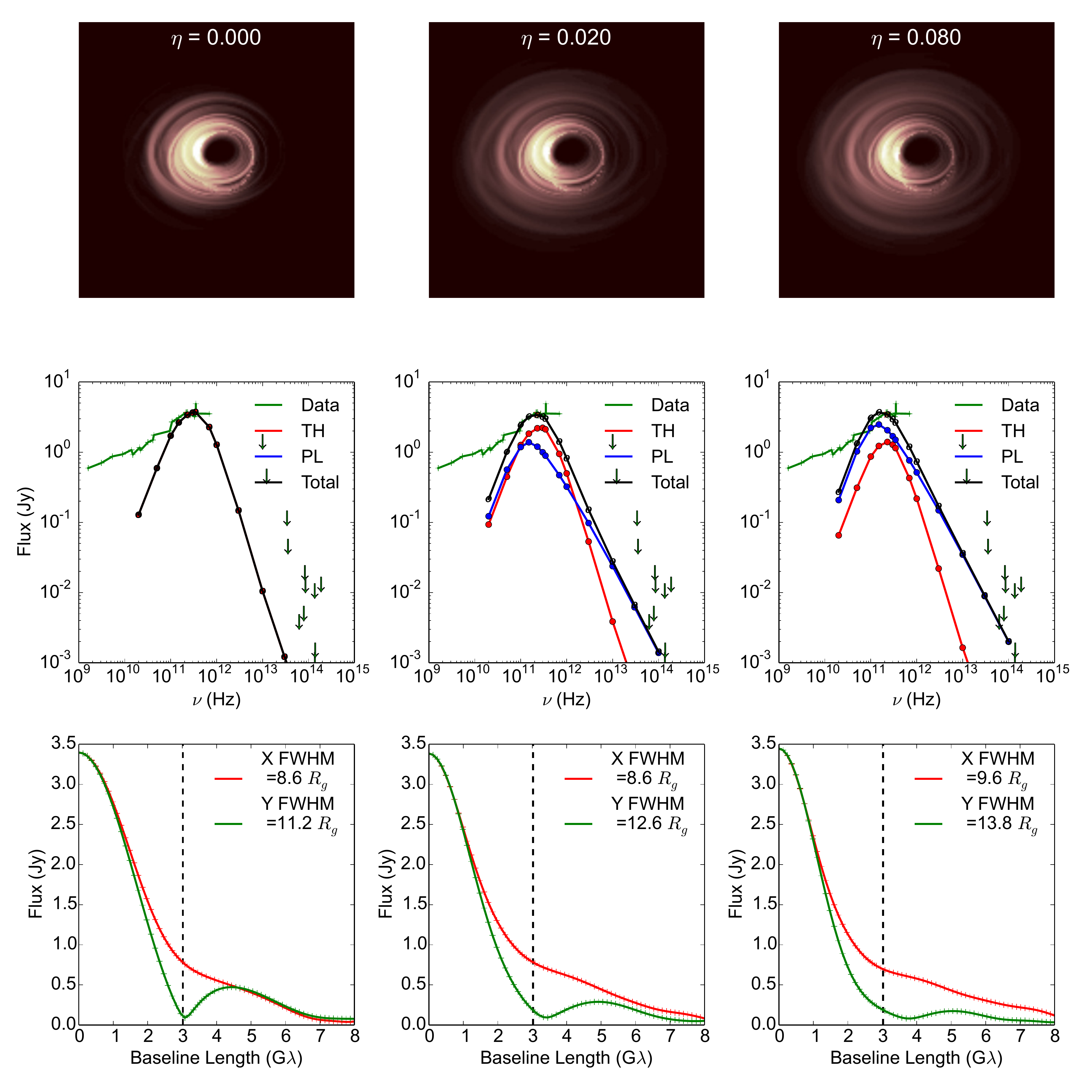}
\caption{\textnormal{As Figure \ref{fig:thermalseq} for hybrid images from purely thermal ($\eta = 0$) to a significant power law tail component ($\eta = 0.08$). For all images \titel{} $= 0.2$.  As the thermal core shrinks, the power law diffuse halo grows with increasing $\eta$. Images depict a 50 x 50 $R_{g}$ region. The total spectra (black, ``Total'') are decomposed into the thermal (red, ``TH'') and power law (blue, ``PL'') components and compared to Sgr A* data (green lines and upper limits as arrows). The data are as described in Figure \ref{fig:thermalseq}. Since the 230 GHz flux is fixed to be 3.4 Jy, the higher $\eta$ images are made with lower number densities.  }
}\label{fig:hybridseq}
\end{figure*}

\begin{figure*}
\includegraphics[width=\textwidth]{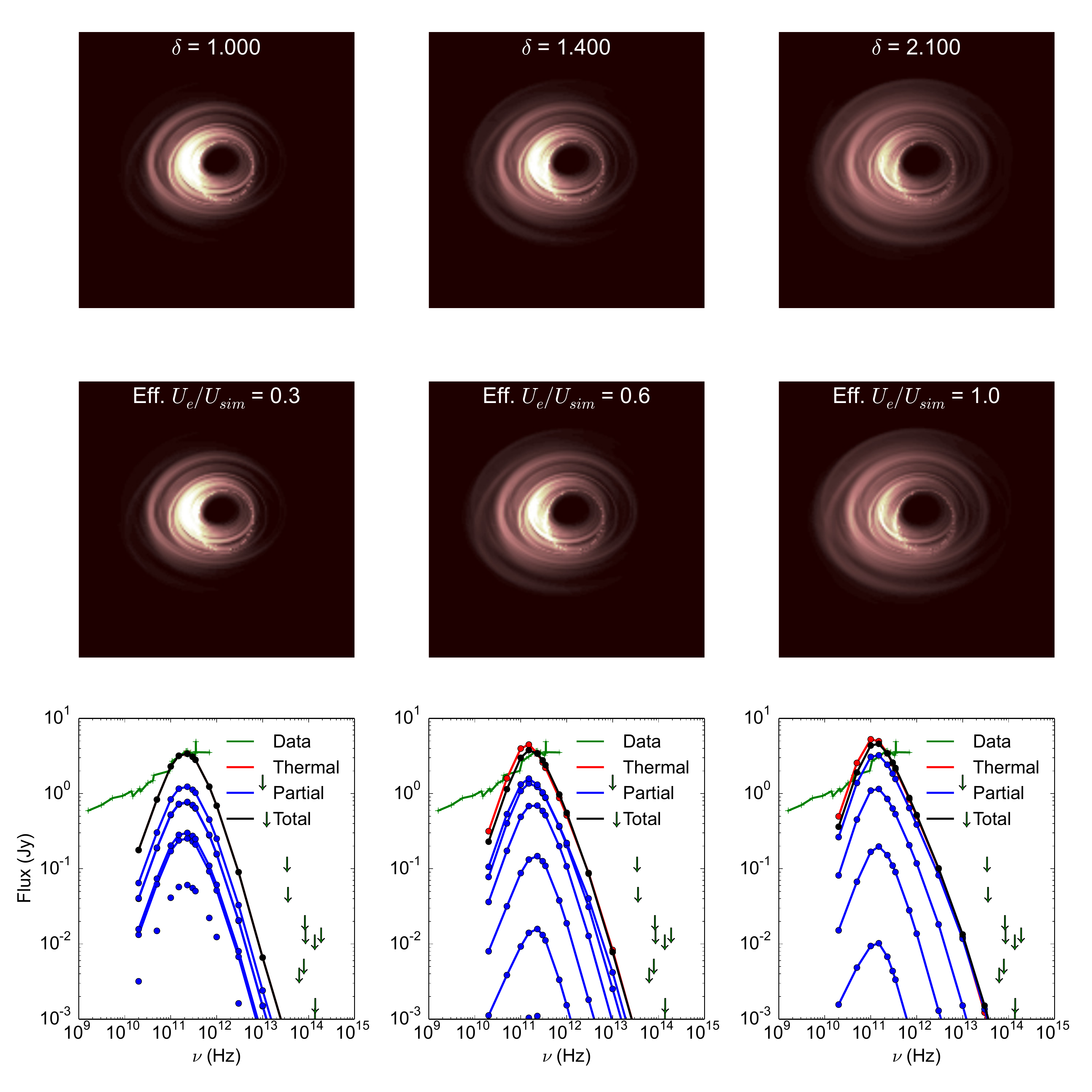}
\caption{A comparison between Fermi acceleration multi-Maxwellian images (top) at fixed total thermal energy (\titel{} $= 0.3$), varying $\delta$ from 1.0 (thermal) to 2.1 \citep[comparable to ][]{lynn2014} and purely thermal images (middle). The thermal images have a temperature corresponding to the highest temperature Maxwellian in the multi-Maxwellian decomposition. The bottom panel compares the total multi-Maxwellian spectra in black to the thermal spectra in red. The spectra of the individual Maxwellian components is shown in blue.  The spectra and images of this single-Maxwellian thermal counterpart are very similar to the multi-Maxwellian model demonstrating that for a sufficiently non-thermal distribution function, lower energy electrons can be completely hidden and not contribute at all to the mm images and spectra.
}\label{fig:maxseq14}
\end{figure*}

\section{Results}\label{sec:results}

To understand the impact of non-thermal electrons on model images of Sgr A*, we compare images from the HARM simulation fluid snapshot at fixed flux for a thermal distribution function with the two non-thermal distributions described in \S \ref{sec:edf}.

\subsection{Thermal distribution}

Figure \ref{fig:thermalseq} shows a sequence of $230$ GHz HARM images (top), spectra (middle), and visibility amplitude profiles (bottom) with fixed inclination $i = 45^\circ$ and increasing electron energy fraction (temperature). The spectra are plotted in comparison with Sgr A* radio \citep{falcke1998,an2005}, submm \citep{marronephd,boweretal2015}, and infrared data. The data points in the mid-infrared are true upper limits \citep{meliafalcke2001}, while we treat the measured near-infrared mean \citep{schoedeletal2011,doddsedenetal2011} and flaring \citep{genzel2003,doddsedenetal2011} flux density measurements as upper limits since their blue spectral shape and high amplitude, short timescale variability characteristics are inconsistent with models like those shown here \citep[e.g.,][]{yuan2003,dexter2010}.

At low electron temperatures, the emission is optically thick and requires a larger emitting area to produce the observed flux at 230 GHz (left panel). With increasing temperature, the photosphere recedes and the image size decreases (middle panel). At still higher electron temperatures (right panel), the image is completely optically thin. Electrons further from the event horizon still have enough energy to significantly contribute to the image, and so the size increases again. At fixed $230$ GHz flux, increasing the temperature then shifts the peak to lower rather than higher frequencies. This is an optical depth effect: at low temperature the flow is optically thick and so the spectrum at $230$ GHz is similar in shape to a blackbody. At high temperature, the accretion flow is optically thin and so the spectral index is steep. This result was previously found in submm models of Sgr A* by \citet{monika2009}.

\textnormal{A prediction of general relativity is that a black hole should cast a shadow with apparent diameter $\simeq 10 R_{g}$, roughly independent of the black hole spin (varying by $\sim 10\%$ with spin) or orientation, due to the location of the circular photon orbit and lensing effects \citep{bardeen1973,falcke,takahashi2004}.} Detecting this feature in mm-VLBI observations of Sgr A* \citep{falcke} is the main science goal of the EHT \citep{doeleman2009whitepaper}. Current mm-VLBI data are too sparse to be used for imaging, and instead models are fit to visibility amplitudes (the absolute value of the Fourier transform of the image). The minima at a baseline length of $\simeq 3 G\lambda$ in the left (both orientations) and middle (green curve) panels of the bottom row in Figure \ref{fig:thermalseq} correspond to the predicted feature of a sharp ring of emission at the location of the photon orbit \citep[e.g.,][modeled as a $\delta$ function of emission and shown as the dashed line]{dexter2009}. \textnormal{In many models, the minimum in the visibility amplitude at $\simeq 3 G\lambda$ corresponds to the presence of bright emission surrounding the shadow \citep[e.g.,][]{bromley2001,broderick2009,dexter2009,dexter2010,monika2009,moscibrodzkafalcke2013,chanetal2015}.}

It is clear from the bottom-right panels of Figures \ref{fig:thermalseq} and \ref{fig:hybridseq}, however, that more complex image structure can move the minimum of the visibility profile away from the simple shadow prediction, or greatly decrease the contrast between the minimum and secondary maximum in the visibility amplitude, even though the shadow is still apparent in the center of the top right image. The presence or lack of the 3 $G\lambda$ visibility minimum then, on its own, does not necessarily confirm or rule out the existence of a black hole shadow in the image. \textnormal{This has also been seen in previous work using thermal distribution functions, in which 
complex image structure and structural variability move the location of the visibility minimum \citep{dexterfragile2013,medeiros2016}. The possible presence of non-thermal electrons adds another layer of uncertainty to its interpretation.}

\subsection{Hybrid Power-law Tail}\label{sec:resultsplt}

Figure \ref{fig:hybridseq} shows a sequence of HARM images from the hybrid distribution function (\S \ref{sec:plt}) with increasing $\eta$ at fixed values of $i = 45^\circ$ and $U_e / U_{sim} = 0.2$. The main effect of changing $\eta$ 
is to increase the fraction of electrons in the power law tail because the fraction $\gamma_{min}/\theta$ remains relatively constant. Since power law tail electrons have values of $\gamma$ typically a factor $\simeq 3-10$ higher than thermal electrons, they radiate more efficiently ($P \propto \gamma^{2}$). Thus, increasing $\eta$ while keeping the 230 GHz flux fixed increases (decreases) the fraction of the total flux produced by power law (thermal) electrons, and decreases the total number density of electrons. This effect will be important in \S\ref{sec:sas}. 

The lower flux produced by thermal electrons causes the thermal ``central core'' of the images in Figure \ref{fig:hybridseq} to shrink with growing $\eta$. The growing fraction of power law flux is contained in a diffuse, low intensity ``halo'' component. The large size of this component is due to the efficient radiation of the high energy power law electrons out to a larger distance from the black hole, and is similar in appearance to images of single temperature Maxwellian models with large \tite{} (see Figure \ref{fig:thermalseq}).

As $\eta$ increases, the overall size of the image becomes larger. Furthermore, the spectrum becomes more like the power law spectrum as the power law contribution to the overall radiated flux increases. The power law tail only requires a small fraction of the total internal energy ($\eta \gtrsim 0.01$) in order to have a noticeable impact on the images and spectra. The spectra found here agree with those from previous work \citep{ozel2000,yuan2003,bro2009b}.

Larger $\eta$ also leads to a steeper 690/230 GHz (submm) spectral index. This is both because the power law component typically peaks at somewhat lower frequency, and because decreasing the flux in the thermal component decreases its optical depth and so steepens its spectrum. Although the example spectrum in Figure \ref{fig:hybridseq} nearly violates a few of the IR limits, this is a result of choosing $p = 3.5$ for the slope of the non-thermal power law tail. Images with a steeper power law (e.g. $p=7$) have less flux at high frequencies, but have similar $230$ GHz images. This is because a small change in $p$ does not alter $\gamma_{min}/\theta$ or $N_{pl}$ by very much, but it does remove the high energy electrons producing near-infrared emission in the model.

The growing image size with increasing $\eta$ is also reflected in the visibility profiles in Figure \ref{fig:hybridseq}. In addition to the overall size increasing, the first minimum in the profile changes in both location and contrast as the power law halo becomes more prominent, even though the black hole shadow is apparent and of the same size in all images. Again this indicates that the location of a minimum in the visibility, if present, depends on the overall image size and shape, and so does not necessarily appear at the predicted location for a surrounding bright ring of emission.

\begin{figure}
\includegraphics[width=\columnwidth]{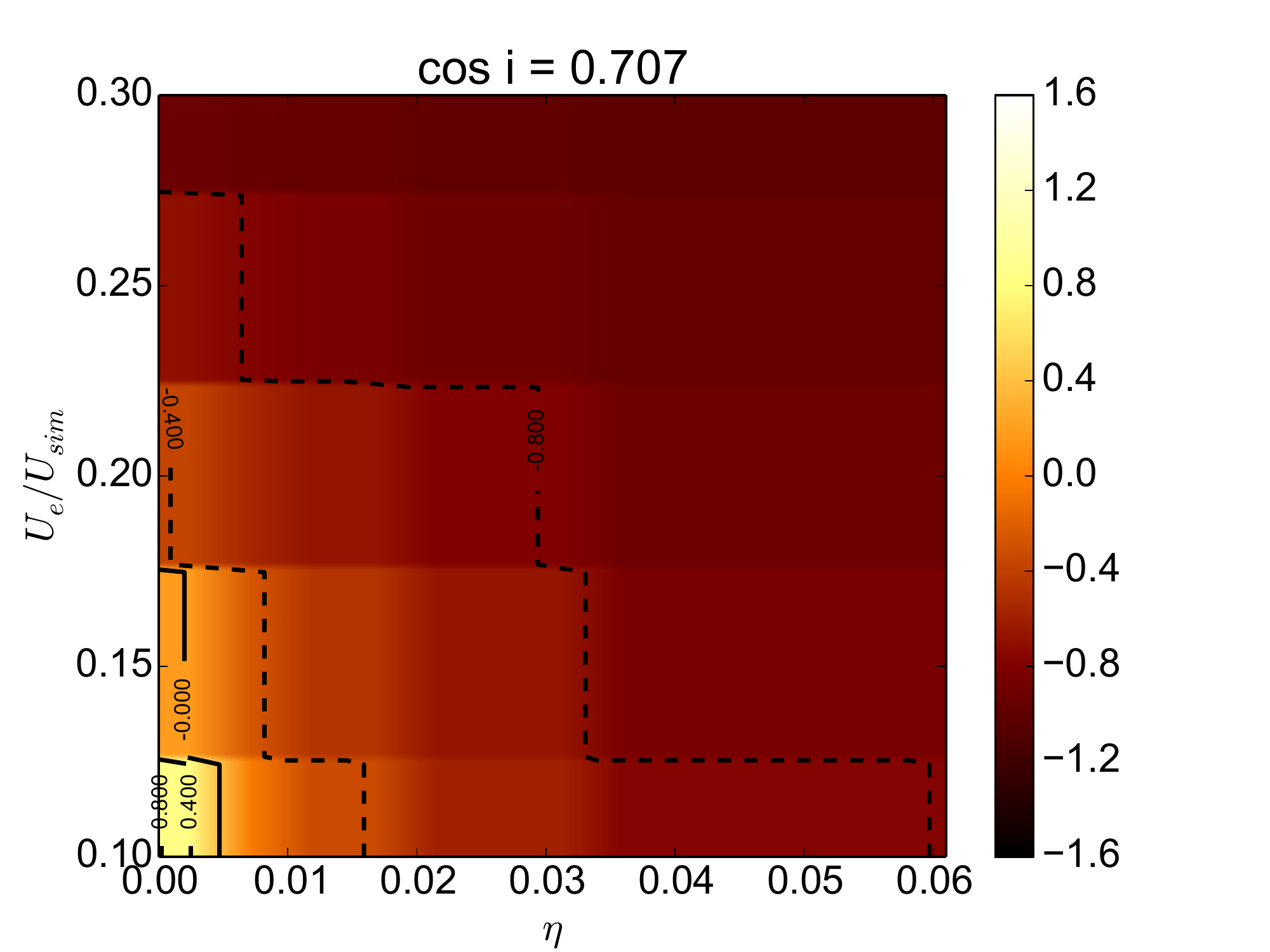}
\caption{A grid of submm 690/230 GHz spectral slopes for Sgr A* calculated from a HARM simulation fluid model with hybrid electron distribution function, varying the power law energy fraction $\eta$ from 0 (thermal) to 0.06 
and the electron energy \titel{} from 0.1 to 0.3. At fixed \titel{}, the spectral index becomes steeper for stronger power law tail components. The observed spectral slope of -0.4 constrains the possible values of $\eta$ (e.g., $\lesssim 0.03$ in this example). 
}
\label{fig:hhcontour}
\end{figure}

Figure \ref{fig:hhcontour} shows a contour plot of $690/230$ GHz spectral index vs. \titel{} and $\eta$. The observed flat submm spectral index $\simeq -0.4$ \citep{marronephd,boweretal2015} constrains the maximum value of $\eta$. This value is degenerate with the total internal energy of the thermal electrons. Although only one inclination is displayed, the maximum value of $\eta$ for other inclinations is still a few per cent. Values of \titel{} below 0.1 require unrealistically high mass accretion rates in order to produce the observed 230 GHz flux, and in these cases $\eta$ must be less than 0.06 to match the observed spectral index. 

\begin{figure}
\includegraphics[width=\columnwidth]{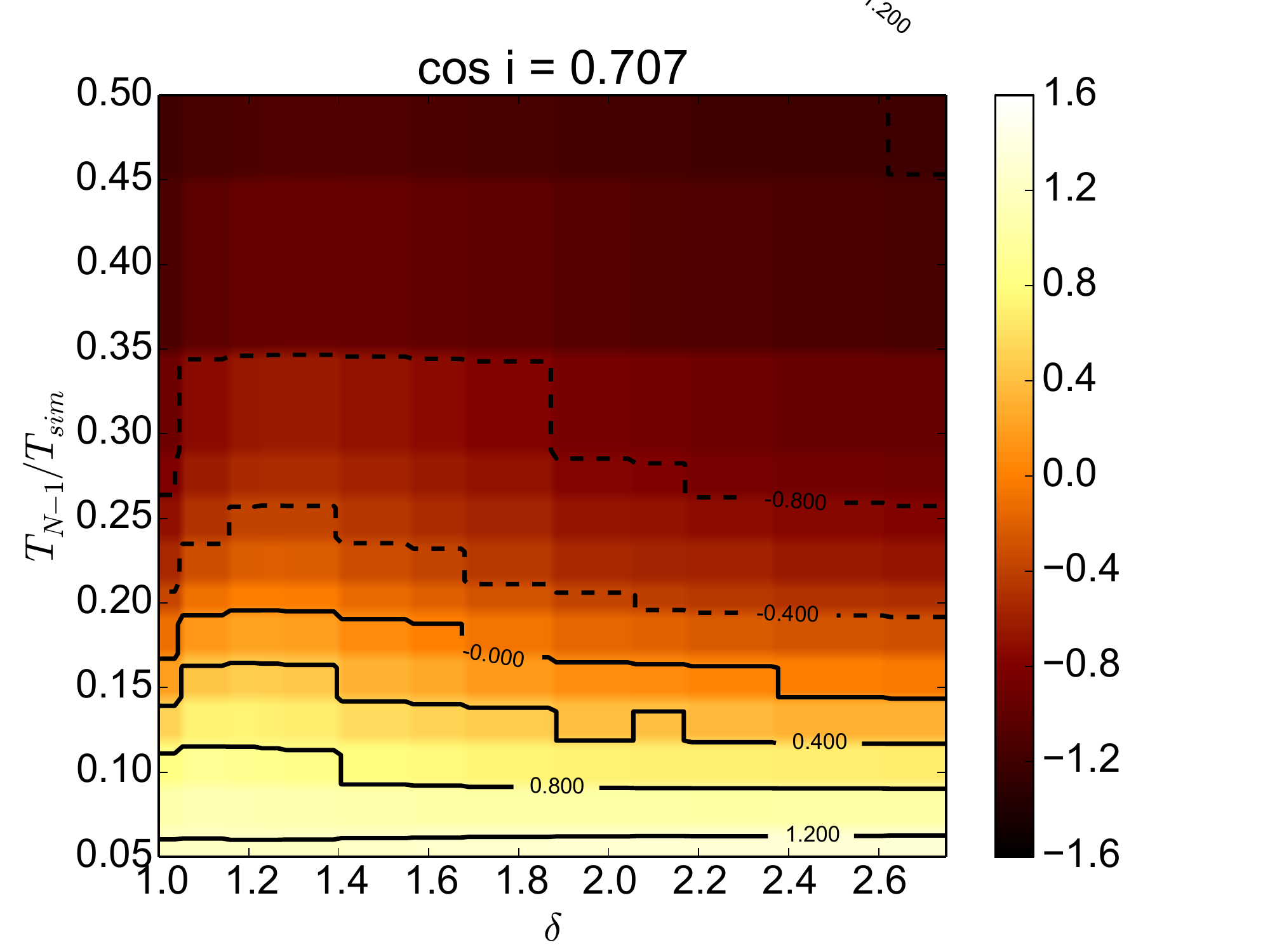}
\caption{A grid of submm 690/230 GHz spectral slopes for Sgr A* calculated from a HARM simulation fluid model with a multi-Maxwellian electron distribution function, varying the temperature spacing $\delta$ from 1.0 (thermal) to 2.8 and the electron temperature of the highest Maxwellian relative to the simulation temperature ($T_{N-1}/T_{sim}$) from 0.05 to 0.5. The spectra are sensitive to the maximum electron temperature, but much less so to the width of the distribution function, since the highest temperature electrons often dominate. The observed spectral slope of -0.4 constrains the temperature of the highest Maxwellian.}
\label{fig:contour}
\end{figure}

\subsection{Multi-Maxwellian Model}\label{sec:resultmmm}

The top two panels of Figure \ref{fig:maxseq14} show a sequence of images and spectra using the HARM simulation fluid model and the Multi-Maxwell distribution function with 
\tite{} = 0.3 and with varying values of the width parameter $\delta$. The spectra are decomposed into each of the six Maxwellian components. For values of $\delta \gtrsim 1.4$ the highest temperature Maxwellian component dominates both the spectrum and the image.  

By using Equation \ref{eq:dmdwsum}, the effective \tite{}
of the highest temperature Maxwellian component can be calculated for each value of $\delta$ by multiplying the overall \tite{} = 0.3 
by $\delta^{N-1}/\displaystyle\sum_{i=0}^{N-1} c_{i}\delta^{i}$. 

We then make equivalent thermal images and spectra (bottom two panels of Figure \ref{fig:maxseq14}) using this effective electron temperature, and varying the total electron number density in order to normalize the flux at 230 GHz. Roughly, since the highest Maxwellian contains $8\%$ of the electrons, the equivalent thermal image number density is about $8\%$ of the total multi-Maxwellian model number density, as expected. However, it is not exact, because the number density is tied to the mass accretion rate and to the magnetic field strength, which affects the emission and absorption coefficients. Nonetheless, it is clear from the spectra in Figure \ref{fig:maxseq14} that the highest temperature Maxwellian causes a small percentage of the electrons to dominate the image and spectrum. 

The agreement between each Multi-Maxwell image and spectrum with its equivalent thermal model is excellent, showing that despite the broad, non-thermal distribution function used in the middle and right panels of Figure \ref{fig:maxseq14}, the results are almost identical to those from using a single temperature distribution function. However, the temperature and number density of this component do not follow in the normal way from the value of \tite{} and $\dot{M}$ chosen: the total number density in the Multi-Maxwell model is much larger, due to the presence of lower energy electrons that do not contribute to the image or spectrum. Similarly, the temperature can be much higher than expected from the value of \tite{} if $\delta$ is large.

Figure \ref{fig:contour} shows the $690/230$ GHz spectral slope as a function of the highest Maxwellian temperature ($T_{N-1}/T_{sim}$) relative to the simulation temperature and temperature spacing $\delta$ in the multi-Maxwellian model. Although only one inclination is displayed, the qualitative dependence of the spectral slope on inclination is similar to that on the temperature: a low temperature is similar to an edge-on inclination. Both amplify the effects of absorption, and edge-on inclinations are predisposed towards larger spectral slopes at higher frequencies. Since the spectra and images are mostly determined by the temperature of the highest Maxwellian component, $\delta$ has little effect on the spectral slope. 

\begin{figure}
\includegraphics[width=\columnwidth]{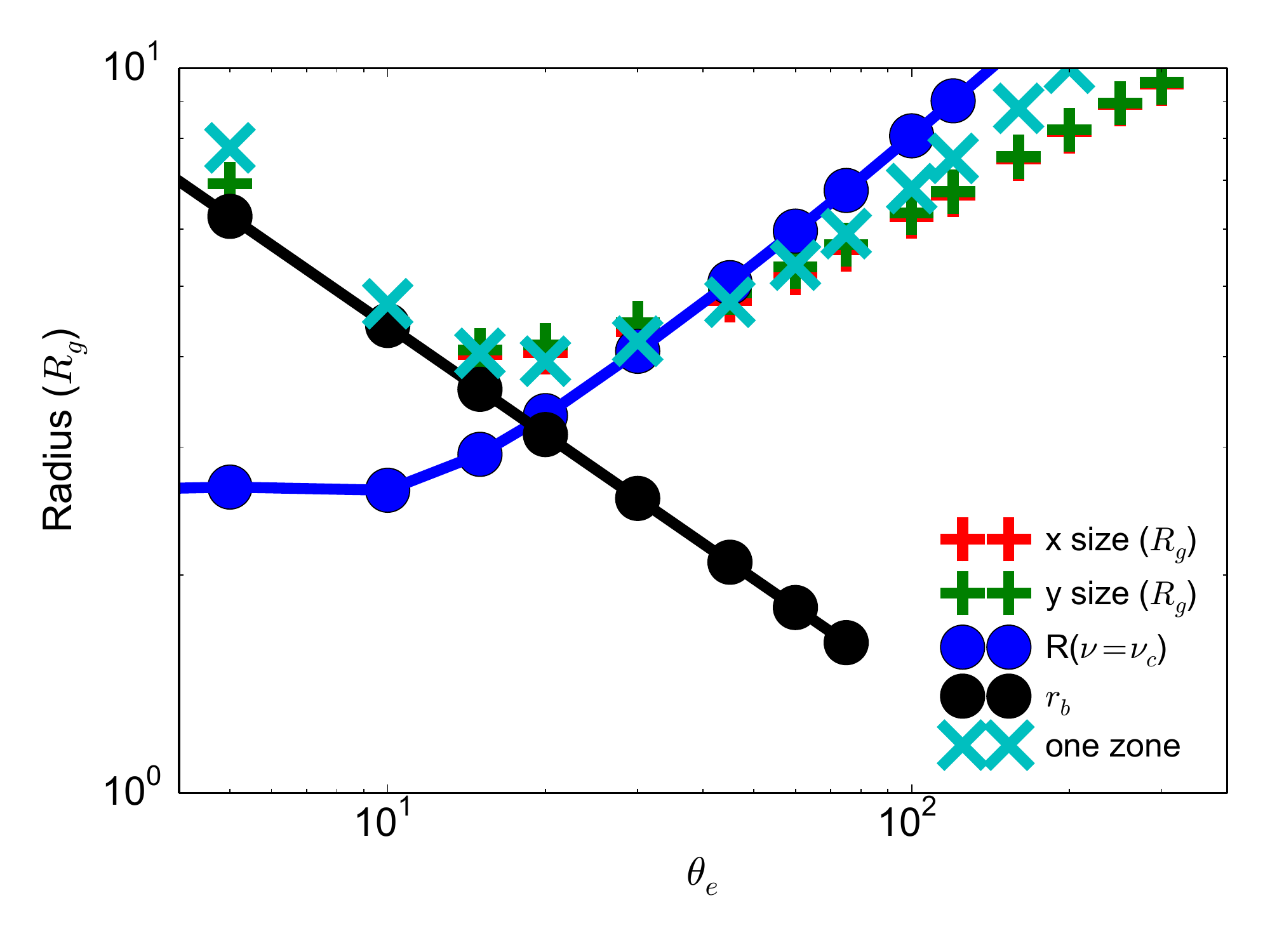}
\caption{A comparison between simulated sizes and estimated analytic sizes. The red ``x size'' and green ``y size'' crosses are sizes derived from relativistic ray tracing, fitting the visibility profile to a Gaussian, and reporting the rms width along two extreme directions. The filled circles are sizes estimated analytically by finding the radius at which the observed frequency is equal to the synchrotron critical frequency (blue), and the radius inferred from equating the brightness temperature to the electron temperature (black), as described in \S\ref{sec:sas}. The cyan ``one zone'' crosses are sizes estimated using a 1-D semi-analytic one zone model, selecting the radius which encapsulates 68\% of the flux. We use a SARIAF fluid model and a thermal electron distribution function. The analytic sizes agree well with the simulated sizes and explain why the size increases with increasing electron temperature. Note that the FWHM sizes would be greater by a factor of $2\sqrt{2\ln{2}}\approx 2.4$. This result generalizes to non-thermal electron distribution functions, replacing the temperature with a typical electron energy.  
}
\label{fig:nunuc}
\end{figure}

\begin{figure}
\includegraphics[width=\columnwidth]{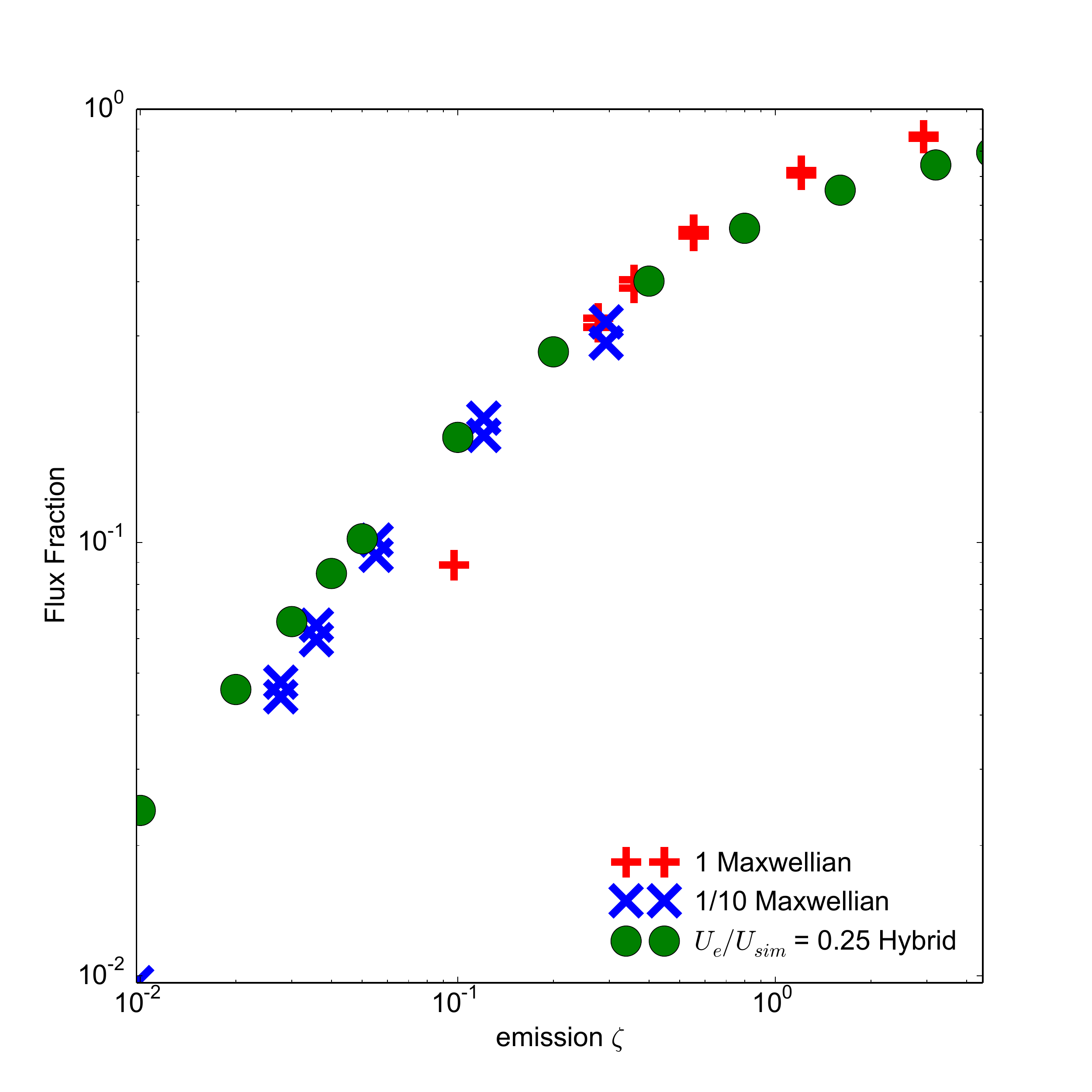}
\caption{Fraction of the total flux in the non-thermal component of various electron energy distributions vs. the emission fraction $\zeta = n \gamma^{2}$ (\eq{zeta}) for the 3 distribution functions considered. ``1 Maxwellian'' is the multi-Maxwellian model with coefficients which fit to the diffusion model of \protect\cite{lynn2014}. ``1/10 Maxwellian'' uses the same coefficients except for the highest Maxwellian, whose coefficient is decreased by a factor of 10 (and all other coefficients are renormalized), which decreases the emission $\zeta$ by a factor of 10. For the multi-Maxwellian models, the flux fraction and emission $\zeta$ shown are those of the highest temperature component. Emission $\zeta$ is adjusted by adjusting the Maxwellian spacing $\delta$, and each Maxwellian model has two points at each value of emission $\zeta$ because we use two values of \titel{}, 0.25 and 0.3. ``0.25 Hybrid'' is the hybrid model described in \S\ref{sec:plt} with \titel{} $= 0.25$, and the flux fraction and emission $\zeta$ of the power-law tail are shown. Despite differences between the 3 distribution functions, emission $\zeta$ is a fairly good predictor of flux fraction. Image sizes for an arbitrary distribution function can thus be estimated by calculating the emission $\zeta$ for its various components. 
}
\label{fig:emeta}
\end{figure}

\section{Analytic interpretation}\label{sec:eis}

For the thermal, power-law tail (\S\ref{sec:plt}), and multi-Maxwellian (\S\ref{sec:dm}) electron energy distribution models, an analytic approach can be used in tandem with a self-similar fluid description (e.g., the SARIAF model, \S\ref{sec:sariaf}) to predict the size of an image. We use these results to understand the numerical image size results from HARM models above, and then show how the results for our two sample non-thermal distribution functions can be generalized to other distribution functions.

\subsection{Image sizes}\label{sec:sas}

We first consider the case of optically thin emission. The spectrum of synchrotron radiation from a single particle of energy $E = \gamma mc^{2}$ has a cutoff on the order of a critical frequency, defined as:

\begin{equation}\label{eq:sasnuc}
\nu_{c} = \frac{3}{4\pi} \gamma^{2} \frac{eB}{m_{e}c}\sin{\alpha}
\end{equation} 

where $e$ is the charge of the electron, $B$ is the magnetic field strength, $m_{e}$ is the mass of the electron, $c$ is the speed of light, and $\alpha$ is the pitch angle between the magnetic field and the particle velocity. 

Since the spectrum drops off sharply for frequencies above $\nu_{c}$, for a given observing frequency $\nu_{o}$, very little emission is expected from regions where $\nu_{c}$ is below $\nu_{o}$. In accretion flows or jets, it is expected that the magnetic field strength $B$ and the internal energy (related to $\gamma$) fall monotonically with increasing distance from the black hole. The critical synchrotron frequency hence also falls monotonically, and the radius at which $\nu_{o} = \nu_{c}$ ($r_{c}$) provides an estimate of the size of the emission region for optically thin plasma. Outside of this radius ($r > r_{c}$), $\nu_{o} > \nu_{c}$, so very little emission is produced. 

In general, for a fluid with magnetic field and temperature that vary as power laws with radius as $B = B_0 \left(\frac{r}{r_{s}}\right)^{-a}$, $\theta(r)=\theta_{e}^{0}\left(\frac{r}{r_{s}}\right)^{-b}$, and with emitting electron energy $\gamma(r) \simeq f \theta(r)$, 

\begin{equation}\label{eq:sasnuc2}
\nu_{c} = \frac{3}{4\pi} (f\theta_{e}^{0})^{2}\left(\frac{r}{r_{s}}\right)^{-2b} B_{0} \left(\frac{r}{r_{s}}\right)^{-a} \frac{e}{m_{e}c}\sin{\alpha}
\end{equation}
and hence, solving for the radius $r_{c}$ at which $\nu_{c} = \nu_{o} = 230$ GHz, we find, 

\begin{equation}
r_{c} = r_{s} \left((f\theta_{e}^{0})^{2} \frac{1}{\nu_{o}}\frac{3}{4\pi}  \frac{eB_{0}}{m_{e}c}\sin{\alpha}\right)^{\frac{1}{a+2b}}.
\end{equation}

As an example for the SARIAF fluid variables, we substitute Equations \ref{eq:sariafn} and \ref{eq:sariafb} into Equation \ref{eq:sasnuc2} with values for $\beta = 10$ and $n_{e,th}^{0} = 10^{7} \mathrm{cm}^{-3}$ to find:

\begin{equation}
\begin{split}
\nu_{c} &= 0.236 \mathrm{GHz}~(\gamma^{2}) \left(\frac{r}{r_{s}}\right)^{-1.05} \\
& \times \left(\frac{\beta}{10}\right)^{-\frac{1}{2}} \left(\frac{n_{e,th}^{0}}{10^{7} \mathrm{cm}^{-3}}\right)^{\frac{1}{2}} \sin{\alpha}.
\end{split}
\end{equation}

\noindent In the hybrid thermal with power-law tail electron energy distribution model, the electrons of interest are those at energies on order of $\gamma_{min}$, where the power-law tail begins. For sufficiently steep power-law slopes, there are not enough electrons at much higher energies for them to contribute significantly. In this case, typically $f \simeq 10$ (see \S\ref{sec:plt}). Using this along with the SARIAF temperature scaling gives a size:

\begin{equation}\label{eq:sasplt}
\begin{split}
r_{c} = 2.3 r_{s} \left(       \left(\frac{f}{10}\frac{\theta_{e}^{0}}{10}\right)^{2}      \left(\frac{\beta}{10}\right)^{-\frac{1}{2}} \left(\frac{n_{e,th}^{0}}{10^{7} \mathrm{cm}^{-3}}\right)^{\frac{1}{2}} \sin{\alpha}\right)^{0.37}
\end{split}
\end{equation}

\noindent at $\nu_{o} = 230$ GHz. 

\begin{figure}
\includegraphics[width=\columnwidth]{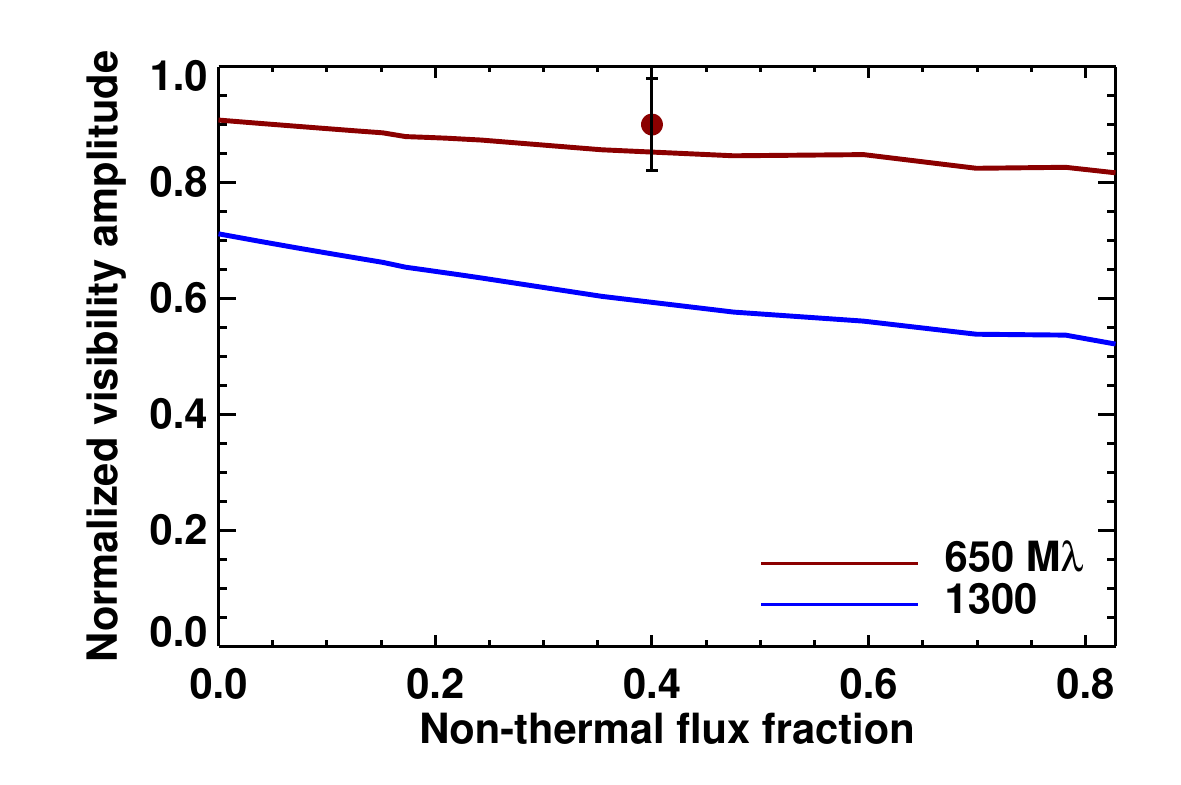}\\
\includegraphics[width=\columnwidth]{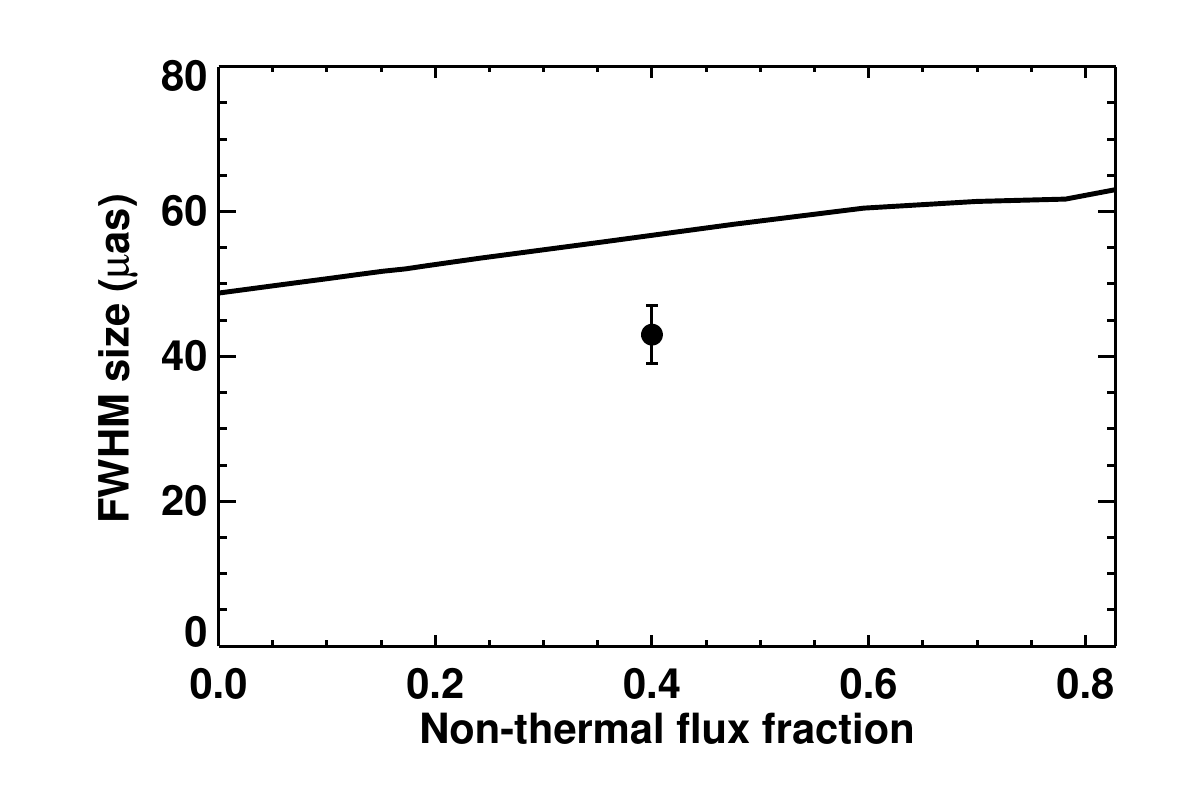}
\caption{The correlated flux density on two characteristic baseline lengths (top) and corresponding best fit Gaussian FWHM sizes (bottom) from the hybrid distribution function and the HARM simulation snapshot as a function of the fraction of the total flux produced by the power law component. The data points show the constraints from current mm-VLBI data \citep{doele2008,fish2011}. Future data, especially on a baseline length corresponding to LMT-SMTO ($\simeq 1300 M\lambda$), should be able to constrain the flux in a non-thermal halo and therefore the electron distribution function.}
\label{fig:ehtsize}
\end{figure}

The power law tail component has $f \simeq \gamma_{min}/\theta_{e}^{0} \simeq 10$. For the thermal component, the typical electron energy $\gamma \simeq 3\theta$ so $f \simeq 3$. From \eq{sasplt}, since the size is proportional to $(f\theta_{e}^{0})^{2/2.73}$, the power law tail component is expected to be $\approx 2-3$ times the size of the thermal component. However, the final image size does not simply grow by this large factor. As discussed in \S\ref{sec:resultsplt}, an increase in $\eta$ at fixed 230 GHz flux necessitates a decrease in the number density. Since the size depends weakly on the number density through its scaling with the magnetic field (\eq{sasplt}) the thermal and power law sizes both decrease. An increase in $\eta$ increases the non-thermal flux fraction, which weights the overall size towards the larger power law size. The shrinking due to number density combined with the growth due to a larger power law halo contribution is consistent with the net increase in size found in the hybrid images (Figure \ref{fig:hybridseq} and Figure \ref{fig:ehtsize}). 

Quantitatively, the image size estimate of \eq{sasplt} also agrees with the most extended part of the image of the high temperature thermal distribution (top right panel of Figure \ref{fig:thermalseq}), and for the diffuse halo in the hybrid case. The field strength and temperature scale differently in the HARM simulation than in the SARIAF model, but are similar enough that using these scalings should not significantly affect the size \citep{dexter2010}. 

The multi-Maxwellian electron energy distribution model is more complicated because the function relating the fluid temperature to the actual electron temperatures for each component ($\theta_{i} = \theta_{0} \delta^{i}$) depends on the coefficients $c_{i}$ and $\delta$ (see \S\ref{sec:dm} and Equation \ref{eq:dmdwsum}, and note $\theta_{0} \neq \theta_{e}^{0}$). The extent to which each Maxwellian component of temperature $\theta_{i}$ contributes also depends on the constants $c_{i}$. However, given $\delta$, $c_{i}$, and assuming a particular Maxwellian component $j$ dominates, the given fluid temperature is multiplied by a constant factor everywhere. At the inner radius $r_{s}$, using $\theta_{j} = \theta_{0} \delta^{j}$ and \eq{dmdwsum},
and assuming $\delta$ and $c_{i}$ are constant, then the Maxwellian temperature that dominates the emission at $r_{s}$ is given by $\theta_{j} = f \theta_{e}^{0}$, where $\theta_{e}^{0}$ is the inner radius fluid electron temperature, and $f$ is now a constant given by:

\begin{equation}\label{eq:sasf}
f = \frac{\delta^{j}}{\displaystyle\sum_{i=0}^{N-1} c_{i}\delta^{i}}.
\end{equation} 

\noindent This value of $f$ can be substituted into Equation \ref{eq:sasplt} in the same manner as the hybrid power-law tail $f$. However, unlike the hybrid power-law tail case, there is no reason to expect $f \approx 10$. 

The estimate in \eq{sasplt} is only valid when the accretion flow is optically thin at the observed frequency $\nu_{o}$. When the flow is optically thick, the intensity is a blackbody from a $\tau \approx 1$ photosphere, so for matching the measured luminosity of Sgr A*, increasing the model temperature necessitates decreasing the size. In contrast, when the accretion flow is optically thin, increasing the model temperature increases the size. In both cases, increasing the model temperature while maintaining the same total flux corresponds to decreasing the particle number density. In the optically thick case, decreasing the particle number density moves the photosphere inward due to decreasing optical depth. In the optically thin case, the size increases because electrons further from the black hole gain enough energy to contribute when the model temperature is increased.

In the optically thick case, an estimate for the image size can be made using the brightness temperature, $T_{b} = \frac{c^{2}}{2\nu^{2}k_{b}} I_{\nu}$ where the brightness is $I_{\nu} = F_{\nu}/\Delta\Omega$ for flux $F_{\nu}$ and solid angle $\Delta\Omega$. For $F_{\nu} = 3.4$ Jy at 230 GHz (1.3 mm), setting the brightness temperature to be equal to the electron temperature $\theta_{e}$, and assuming a circular source with angular radius $\Delta\phi$, the angular size is thus estimated as,

\begin{equation}
\Delta\phi = 69 ~ \rm \mu as~\theta_{e}^{-1/2}.
\end{equation}

For the gravitational radius of Sgr A* $r_{g} = G M / c^2 \approx 6 \times 10^{11}$ cm and the distance 8 kpc, the image size estimate from the brightness temperature ($r_{b}$) thus becomes,

\begin{equation}\label{eq:sasbright}
r_{b} = 7~r_{s}~\theta_{e}^{-1/2} .
\end{equation}

\textnormal{In the hybrid case, \cite{ozel2000} showed that the thermal electrons provide the absorption at low temperatures while the non-thermal electrons provide the emission. A more sophisticated version of the brightness temperature estimate would account for the change in source function, but otherwise the same argument holds.}

We next quantitatively compare these analytic size estimates to numerical ones using simulated 2-D images with a SARIAF fluid model (\S\ref{sec:sariaf}) and a thermal electron distribution function (\S\ref{sec:dm}). In general, the larger of the two estimates $r_{b}$ and $r_{c}$ provides a good predictor of the size. The face-on ($\cos{i} \approx 1$) case was chosen because it reduces effects of asymmetry, Doppler beaming, and absorption by the disk. The size was measured for these images by fitting a Gaussian to the Fourier transform and taking the standard deviation (not the FWHM). We also compare the ray-tracing simulation results with sizes taken from a 1-D semi-analytic one zone model. Using the SARIAF fluid model and synchrotron emission and absorption coefficients, we produce radial brightness profiles. The radius which encloses 68\% of the flux is taken to be the size estimate. The result is shown in Figure \ref{fig:nunuc}, which demonstrates the behavior described in the analytics above. Not only does the size first fall then rise with increasing temperature, but also there is good quantitative agreement between the measured sizes (x size and y size) and the predicted sizes (critical frequency radius $r_{c} = r(\nu=\nu_{c})$ and brightness temperature radius $r_{b}$). This is especially true at the low temperature end (very optically thick) and high temperature end (very optically thin), where the analytic approximations are most applicable. 

\subsection{Emission Fraction}\label{sec:emeta}

Although images of the hybrid power-law tail and multi-Maxwellian models at first appear to be completely different, this is partially because most examples of the power-law tail have a non-thermal component which contains a small percentage of the total thermal energy (a few per cent), whereas our examples of the multi-Maxwellian models have high temperature components which have a large fraction of the total thermal energy (tens of per cent). In order to examine this effect, we define an emission fraction $\zeta$ as:

\begin{equation}\label{eq:zeta}
\zeta = \frac{n_{a}\theta_{e,a}^{2}}{n_{b}\theta_{e,b}^{2}},
\end{equation}

where $n_{a}$ is the number density and $\theta_{e,a}$ is a typical electron energy of component ``a'' (here, the non-thermal component) and similarly for component ``b'' (here, the thermal component). This is motivated by approximating the emission to be dominated by a single-energy population of electrons.

For the hybrid power-law tail, the energy density is proportional to $n_{th}\theta_{e}$ for the thermal component and $n_{pl}\gamma_{min}$ for the power-law component. Since the energy density fraction is $\eta$ and $\gamma_{min} \approx 10\theta_{e}$, emission $\zeta = \frac{n_{pl}\gamma_{min}^{2}}{n_{th}\theta_{e}^{2}} = \frac{\gamma_{min}}{\theta_{e}}\eta \approx 10\eta$. 

For the multi-Maxwellian case, the number density ratios can be replaced with coefficient ratios, and the temperature ratios can be replaced with powers of $\delta$. Taking the $c_{5}$ component with temperature $\theta_{0}\delta^{5}$ (the highest temperature Maxwellian) to be the non-thermal component and the sum of the other components to be the thermal component, the equation for emission fraction $\zeta$ for the multi-Maxwellian case becomes:

\begin{equation}
\zeta = \frac{c_{5}(\delta^{5})^{2}}{\displaystyle\sum_{i=0}^{4} c_{i}(\delta^{i})^{2}}
\end{equation}
\noindent Figure \ref{fig:emeta} illustrates a relationship between the emission fraction $\zeta$ and the flux fraction (non-thermal total flux divided by combined total flux). We use 3 distribution functions with a similar value of \tite{} $\sim 0.3$. We adjust the emission fraction $\zeta$ by adjusting $\delta$ for the multi-Maxwellian models and $\eta$ for the Hybrid thermal and power law model.  
Figure \ref{fig:emeta} suggests a rough method to predict the flux fraction for components of a given non-thermal distribution, using the emission fraction $\zeta$. It is particularly striking that results for 3 different distribution functions follow a single relation.  

For a general electron distribution function decomposed into Maxwellian and power-law components, the emission fraction provides a quick way to estimate which of the components will dominate or contribute to the emission. In the specific case of the multi-Maxwellian model with coefficients $c_{i}$, by comparing emission $\zeta_{i} = c_{i}\delta^{2i}$ for a given $\delta$, it is clear why the highest temperature Maxwellian dominates for $\delta \gtrsim 2$. The lowest temperature components are suppressed by powers of $\delta^{2}$ and the second highest Maxwellian only has a factor of $\sim 3$ larger $c_{i}$. In the specific case of the hybrid power-law tail model, since emission $\zeta \approx 10\eta$, even a small $\eta \approx 0.02$ power-law component will contribute significantly, and $\eta \approx 0.1$ will dominate, which is apparent in Figure \ref{fig:hybridseq}.
A general electron distribution function can be decomposed into Maxwellian and power law components. Each component can be assigned a relative emission $\zeta$ using \eq{zeta}. When $\zeta$ is large for a single component, that single component will dominate, as in the multi-Maxwellian model. When $\zeta$ is large for multiple components, multiple components can be expected to contribute significantly. After identifying the components of the distribution function that contribute significantly to the emission, 
the image size can be roughly estimated using the approach of \S\ref{sec:sas} (in particular by calculating $r_{b}$ and $r_{c}$ for the relevant components). In cases where 1 component dominates, many properties of the image and spectrum can be estimated from a single component, ignoring the other components. 

\section{Discussion}

We have used two simple, physically motivated examples to study the effect of the electron distribution function on resolved images of synchrotron emission from low-luminosity black hole accretion flows, with the particular application to EHT observations of Sgr A*. The intensity profile (and so observed image size) depends on the energy of the emitting electrons (\S\ref{sec:sas} and Figure \ref{fig:thermalseq}). Cold electrons (with $T_e < T_b$ for brightness temperature $T_b$) require a large emitting area to produce the observed flux density ($\simeq 3$ Jy for Sgr A* at $230$ GHz), and therefore a large image size, which grows with decreasing temperature at fixed flux density. The image size reaches a minimum as $T_e \approx T_b$ and the fluid becomes optically thin down to the black hole. For $T_e > T_b$, a range of radii contribute significantly to the intensity profile, out to a radius $r(\nu=\nu_c)$, where $\nu_c$ is the critical frequency for synchrotron radiation. This frequency scales as $\nu_c \propto T_e^2$, and so the maximum radius and image size increase with increasing temperature. 

A high energy power law tail of electrons can produce an extended image, whose total flux is non-negligible even when they contain a small fraction (few per cent) of the total internal energy density and even when the non-thermal electrons do not substantially change the mm spectrum. The size of this halo is set by the radius where $\nu \simeq \nu_c$. The size of an observed image constrains the amount of flux density arising from any given size, which in turns constrains the distribution function and magnetic field strength profile. 

Figure \ref{fig:ehtsize} shows an example of this for the single HARM snapshot with fixed parameters considered here. The top panel shows the correlated flux density on two baselines as a function of the fraction of flux in the power law component, which corresponds to $\eta$ in our hybrid model. For increasing amounts of flux in the power law, the correlated flux on shorter baselines drops due to resolving out the more extended halo component. Similarly, the best fit FWHM Gaussian size increases (bottom panel). The magnitude of the effect is comparable to the errors in current data, and so should be possible to constrain with future mm-VLBI data. 

A broadened thermal distribution, motivated by a Fermi acceleration model for electron energy evolution \citep{lynn2014}, instead leads to images that are dominated by the highest temperature electrons. This is because the high energy electrons in this case receive a significant fraction of the internal energy, $\eta \gtrsim 0.1$, and so dominate the emission. The bulk of the particle distribution (at lower energies) contributes negligibly to the emission in this case. Fitting the spectrum to data only constrains the properties of the emitting particles. For this reason, estimates of the accretion rate from the emitting particle density \citep{monika2009,dexter2010,shcherbakovetal2012} are lower limits, since they do not constrain the properties of colder electrons. Similarly, the image size is only sensitive to the typical electron energies of the hottest subpopulation of electrons. Faraday rotation can independently constrain the properties of the colder electrons that do not produce significant emission. 

The electron distribution function also affects the shape of the peak of the observed spectrum. At fixed flux density, higher energy electrons have \emph{softer} spectra, since lower particle densities and therefore optical depths are required to produce the observed flux. The spectral index therefore becomes more negative with increasing $\eta$, mimicking the effect of a lower black hole spin or observer's inclination angle \citep[e.g.,][]{monika2009}. For a given observed spectrum then, using models with $\eta = 0$ could bias estimated parameters towards lower black hole spin or inclination if the true distribution is non-thermal.

For a thermal electron distribution function with high electron temperature (right panel of Figure \ref{fig:thermalseq}), a multi-Maxwellian distribution function with a high temperature component (middle and right panels of Figure \ref{fig:maxseq14}), or a hybrid electron distribution with a power-law tail (Figure \ref{fig:hybridseq}), the presence of high energy electrons leads to a larger, more diffuse image. 
The prediction of a minimum of the visibility profile at $3G\lambda$ is the first minimum of the square of a Bessel function $J_{0}$, the Fourier transform of a thin ring of emission around the black hole shadow. However, with a larger, more diffuse image, additional emission is produced at larger radii, changing the Fourier transform from a simple Bessel function. 
This change in the image structure moves the first minimum of the visibility profile away from the expected shadow prediction at $3G\lambda$ and decreases its contrast. However, the black hole shadow is clearly visible in our simulated images. Hence, the location of the first minimum in the visibility profile does not directly reflect the size of the black hole shadow. Uncertainties in the electron distribution function around Sgr A* need to be considered for interpreting EHT observations, particularly given that a small non-thermal component can significantly change the visibility profiles. 

The hybrid particle distribution function used here is taken from previous work with semi-analytic RIAFs \citep{ozel2000,yuan2003}. \citet{bro2009b} calculated images from a similar model with a hybrid distribution function, but did not find the same extended ``halo'' we find here. This is because \citet{bro2009b} use a much more compact spatial distribution for their non-thermal electrons than for thermal ones, preventing the halo from forming. 

\textnormal{We expect the results of this study to generalize to images and spectra from an arbitrary distribution function and to a variety of accretion disk models. The results of \S\ref{sec:eis} do not assume a particular form of the distribution function and can be applied to any underlying fluid model approximated with power laws of radius.}

\textnormal{
In this paper, we have also investigated a method of decomposing arbitrary distribution functions into Maxwellian and power law components, which is not only convenient for radiative transfer calculations (reducing $\simeq 100$ integration steps to $\simeq 10$ components) but also for understanding the contribution of the different components ($\simeq 1-3$ dominant components) of the distribution function to the observed emission.}

\textnormal{This method provides an alternative to choosing a particular form for a non-thermal distribution function \citep[e.g.,][]{pandyaetal2016}, is implemented in the public code \textsc{grtrans} \citep{dexter2016}, and would be straightforward to use with any other radiative transfer code. Moreover, it provides a quick way to make analytic estimates from an arbitrary distribution function, even if the latter lacks a tractable analytic form.}

\section*{Acknowledgments}
We thank the referee, Feryal \"{O}zel, for thorough and constructive comments which improved the paper. This work was supported in part by NSF grant AST 13-33612, a Simons Investigator Award from the Simons Foundation, the David and Lucile Packard Foundation, a Sofja Kovalevskaja Award from the Alexander von Humboldt Foundation of Germany, and a UC-HiPACC Summer Astrocomputing Project Grant. This material is based upon work supported by the National Science Foundation Graduate Research Fellowship under Grant No. DGE-1148900.

\bibliographystyle{mnras}
\bibliography{combine}

\bsp
\label{lastpage}
\end{document}